\documentclass[aps,reprint,prb,twocolumn,floatfix,amsmath,amssymb]{revtex4-2}

\usepackage{graphicx}
\usepackage{dcolumn}
\usepackage{bm}
\usepackage{algorithm}
\usepackage{algorithmic}
\usepackage{colortbl}
\definecolor{lightblue}{RGB}{173, 216, 230}
\definecolor{lightgreen}{RGB}{144, 238, 144}

\begin{document}

\preprint{APS/123-QED}

\title{Photonics-Enhanced Graph Convolutional Networks}

\author{Yuan Wang}
\affiliation{School of Mathematical and Physical Sciences, University of Sheffield, Sheffield S10 2TN, United Kingdom}

\author{Oleksandr Kyriienko}
\affiliation{School of Mathematical and Physical Sciences, University of Sheffield, Sheffield S10 2TN, United Kingdom}


\begin{abstract}
Photonics can offer a hardware-native route for machine learning (ML). However, efficient deployment of photonics-enhanced ML requires hybrid workflows that integrate optical processing with conventional CPU/GPU based neural network architectures. Here, we propose such a workflow that combines photonic positional embeddings (PEs) with advanced graph ML models. We introduce a photonics-based method that augments graph convolutional networks (GCNs) with PEs derived from light propagation on synthetic frequency lattices whose couplings match the input graph. We simulate propagation and readout to obtain internode intensity correlation matrices, which are used as PEs in GCNs to provide global structural information. Evaluated on Long Range Graph Benchmark molecular datasets, the method outperforms baseline GCNs with Laplacian based PEs, achieving $6.3\%$ lower mean absolute error for regression and $2.3\%$ higher average precision for classification tasks using a two-layer GCN as a baseline. When implemented in high repetition rate photonic hardware, correlation measurements can enable fast feature generation by bypassing digital simulation of PEs. Our results show that photonic PEs improve GCN performance and support optical acceleration of graph ML.
\end{abstract}

\maketitle

\section{Introduction}

The use of photonic systems for solving machine learning (ML) tasks has gained significant attention \cite{Shastri2021} and recently seen major breakthroughs \cite{Hua2025,Ahmed2025,Kalinin2025}. These include demonstrations of optical and photonic neural networks achieving ultrafast matrix multiplications~\cite{LinOzcan2018,Hamerly2019,zhou2022photonic,Ma2025}, photonic reservoir computing~\cite{vandoorne2014experimental,Larger2017,Butschek2022,wang2024ultrafast}, and large-scale integrated photonic accelerators for linear algebra~\cite{Feldmann2021,Hua2025,Ahmed2025,Xu2024Taichi}. The inherent parallelism and low latency properties of optical systems \cite{McMahon2023} make them particularly attractive for computationally intensive ML workloads, with orders of magnitude improvements in energy efficiency compared to electronic counterparts~\cite{wang2022optical,wright2022optonn,Ma2025}. Latest advances also show that optical networks can be made reconfigurable \cite{Onodera2025}, extended to programmable nonlinear operation \cite{Yanagimoto2025}, and realized in exciton-polariton platforms for neuromorphic and nonlinear computing \cite{Ballarini2020,Mirek2021Nano,Opala2025,Matuszewski2021,Opala2023}. Finally, photonic neural networks operated in the quantum regime offer access to computational speedups via quantum interference and measurement statistics in multiphoton photonic circuits \cite{Steinbrecher2019,Cimini2024,Zia2025,Yin2025,Monbroussou2025,Ghosh2019,Ghosh2021,verstraelen2025,ShangYu2025,rambach2025}.

To date, most photonic ML demonstrations have focused on classification tasks, including handwritten digit and image recognition \cite{LinOzcan2018,Ma2025,Feldmann2021}, sensing \cite{wang2023image_sensing_onn,Cimini2024}, speech recognition \cite{vandoorne2014experimental,Larger2017,wang2024ultrafast}, and state discrimination \cite{Yin2025,Zia2025,Ghosh2021}. However, extending photonic ML to graphs and relational data remains challenging and comparatively less mature. While early photonics-based~\cite{Yan2022DGNN,Tang2022OGNN} and polaritonics-based~\cite{yuanwang2025} approaches to graph processing started to emerge, hybrid schemes that combine the strengths of photonic operation with the representational power of standard graph neural networks (GNNs) are still largely unexplored.

As standalone solutions, GNNs are widely used for learning on non-Euclidean data \cite{Bronstein2021GGGGG}, with applications in molecular chemistry~\cite{duvenaud2015convolutional,gilmer2017neural}, social networks~\cite{kipf2016semi}, and physics simulations~\cite{sanchez2020learning}. Among different GNN architectures, graph \emph{convolutional} networks (GCNs) in particular primarily aggregate information from local neighborhoods. Despite their strong empirical performance, GCNs can miss important global structural patterns and face fundamental limitations when modeling large molecular graphs~\cite{alon2020bottleneck,topping2021understanding}. In particular, over-smoothing~\cite{li2018deeper,oono2019graph} and over-squashing~\cite{alon2020bottleneck,topping2021understanding,DiGiovanni2024} hinder the propagation of information beyond local neighborhoods, limiting the ability to capture long-range molecular interactions that are critical for property prediction~\cite{dwivedi2022long}. It was shown that equipping GNNs with positional embeddings (PEs) can partially address these limitations~\cite{dwivedi2023benchmarking,kreuzer2021rethinking}, leading to improved performance. While Laplacian eigenvector-based positional embeddings (LapPE) achieved great success~\cite{belkin2003laplacian,dwivedi2022long}, computing these PEs is time demanding, and often they may not capture the complex interaction patterns present in molecular systems. Recent works have suggested positional encodings for graph learning based on quantum evolution and correlation readout in programmable neutral-atom systems \cite{Thabet2024,Faria2025}. This raises two questions: how much advantage can coherent walks on graphs provide, and how can their physical implementation translate into absolute runtime gains?
\begin{figure*}[t]
\begin{center}
\includegraphics[width=1.0\linewidth]{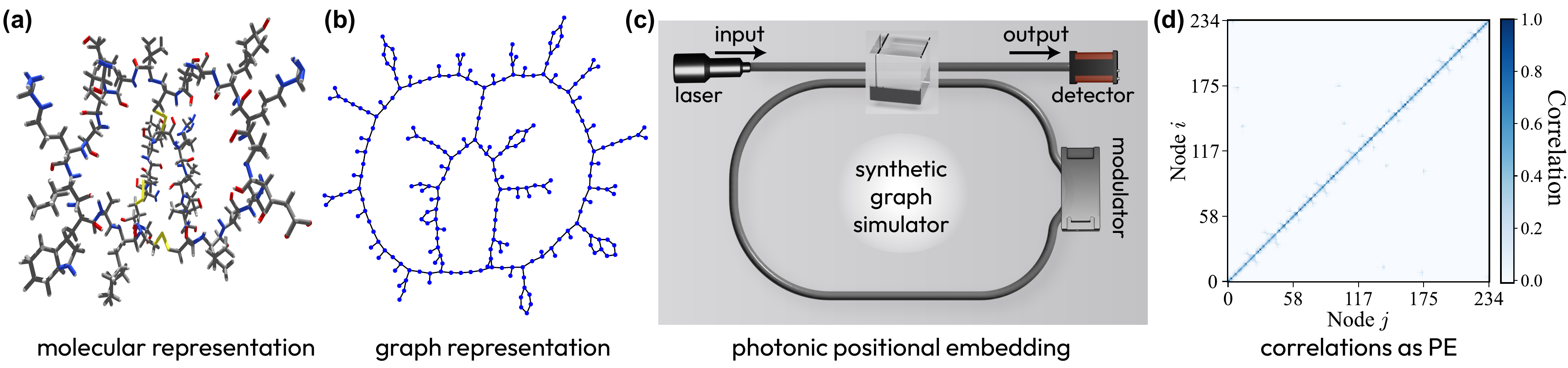}
\end{center}
\vspace{-2\baselineskip}
\caption{\textbf{Workflow for photonics-enhanced machine learning of molecular graph structures using synthetic frequency lattices.} (a) Chemical structure of a peptide molecule ($\mathrm{C_{148}H_{319}N_{41}O_{40}S_{6}}$) representing a sample from the \texttt{Peptides-struct} and \texttt{Peptides-func} datasets used for graph convolutional network analysis. (b) Graph representation $\Gamma$(Peptide) of the molecular structure shown in (a), where nodes represent heavy atoms and edges represent chemical bonds, providing the connectivity information for network analysis. (c) Schematic of the programmable photonic simulator employing synthetic frequency lattices (c.f. Ref.~\cite{senanian2023programmable}), where the intracavity field evolution enables mapping of arbitrary graph structures through phase modulation. (d) Normalized intensity-intensity correlations $\{G_{ij} \}$ for nodes $i$ and $j$, obtained from the photonic simulation setup in (c).}
\label{fig:workflow}
\end{figure*}

In this work, we address these questions and enhance GCNs by introducing a photonics-based PE, where light propagation on graph-shaped lattices is used for feature engineering (Fig.~\ref{fig:workflow}). Our approach is motivated by realizations of optical synthetic frequency lattices~\cite{Yang2022,Yuan2018,senanian2023programmable}, which can be programmed into graph-like coupling structures and support low-latency readout of correlation functions \cite{senanian2023programmable}. Experimental platforms can reach programmable lattices with over $100{,}000$ sites, providing a scalable route to extracting structure-sensitive features at high sample processing rates for graph ML workflows. In contrast to purely mathematical constructions based on spectral decomposition (like LapPE \cite{belkin2003laplacian,dwivedi2023benchmarking}), we propose to use response of graph-shaped synthetic lattices, where the underlying physics naturally encodes long-range couplings and correlations that are central to challenging graph tasks~\cite{dwivedi2022long}. We show that photonic embeddings yield performance gains on molecular datasets, and outline how optical accelerators can reduce time-to-solution by enabling fast feature generation. More broadly, our works points to a direction in which advances in programmable photonic systems can interface effectively with ML workflows.


\section{Model}
\label{model}

We propose a workflow for physics-informed graph analysis, and study graph regression and classification within three approaches based on: 1) GCNs as the baseline architecture; 2) LapPE extracted from the spectrum of the normalized Laplacian; 3) and photonic positional embeddings (PhotPE) to capture global graph structure. The analysis is performed for molecular datasets with complex long-range dependencies, specifically targeting peptide structures. Below, we describe each component of our workflow.

\textit{Graph Convolutional Networks.---}GCNs form the foundation of our approach, performing local message passing through graph convolutions. Let $G = (V, E)$ denote an undirected graph with $n$ vertices, where $V$ is the vertex set and $E$ is the edge set. We denote its weighted adjacency matrix as $A \in \mathbb{R}^{n \times n}$, where $A_{ij} = A_{ji} > 0$ if and only if $(v_i, v_j) \in E$, and zero otherwise. The degree matrix is $D = \mathrm{diag}(d_1, \ldots, d_n)$ with $d_i = \sum_j A_{ij}$. A singe GCN layer~\cite{kipf2016semi} updates node features $X \in \mathbb{R}^{n \times d}$ according to the map 
$H = \sigma\left(\widetilde{D}^{-1/2} \widetilde{A} \widetilde{D}^{-1/2} X \Theta\right)$, where $\widetilde{A} = A + \mathbb{I}$ and $\widetilde{D} = D + \mathbb{I}$ include self-loops, where $\Theta$ is a learnable weight matrix, $\mathbb{I}$ is an identity matrix, and $\sigma$ is a nonlinear activation function. This operation aggregates features from immediate neighborhood of each node.

\textit{Laplacian Positional Embeddings.---}The standard approach to incorporating global structural information in GNNs employs eigenvectors of the graph Laplacian \cite{dwivedi2023benchmarking,Kreuzer2021}. The symmetric normalized Laplacian $L_{\text{sym}} = \mathbb{I} - D^{-1/2} A D^{-1/2}$ encodes the spectral properties of the graph, with its eigendecomposition $L_{\text{sym}} = U_L \Lambda_L U_L^{\top} $ providing a basis for graph Fourier analysis. The first $k$ non-trivial eigenvectors from $U_L$, corresponding to the smallest non-zero eigenvalues, capture smooth variations across the graph and serve as positional features. These Laplacian eigenvectors $U_{L,k} = [u_{L,2} \| u_{L,3} \| \cdots \| u_{L,k+1}]$ are normalized to zero mean and unit variance for stability across different graphs, then projected and concatenated with node features: $\mathcal{P}_L = [X \Vert W_{\text{proj}}\widetilde{U}_{L,k}]$. While LapPE have proven effective for improving GNN expressivity and have become a standard baseline \cite{dwivedi2023benchmarking}, they represent purely mathematical constructs derived from spectral graph theory, lacking physical interpretability for molecular systems where actual wave propagation and interaction dynamics play crucial roles. Thus, the injection of the LapPE for each GCN layer is performed by concatenation $\widetilde{H}^{(\ell)} = [H^{(\ell)} \Vert \mathcal{P}_L]$. Further details of LapPE implementation are provided in Methods section.
\begin{table*}[bt]
\caption{\label{tab:results_main}Baselines for \texttt{Peptides-struct} and \texttt{Peptides-func} datasets~\cite{dwivedi2022long}. Performance metric is mean absolute error (MAE; lower is better) and average precision (AP; higher is better) for \texttt{Peptides-struct} and \texttt{Peptides-func} tasks, respectively. Each experiment was run with $4$ different seeds with $2$ and $3$ GCN layers. N/A: Not applicable. \textbf{Bold}: Best score.}
\begin{ruledtabular}
\begin{tabular}{c c c c c c c}
& &
 & \multicolumn{2}{c}{\texttt{Peptides-struct} (Regression)} & \multicolumn{2}{c}{\texttt{Peptides-func} (Classification)}
\\
\textrm{Model} & 
\# \textrm{Layers}&
\# \textrm{Dim. PE}&\# \textrm{Parameters} & Test MAE ↓
& \# \textrm{Parameters} & Test AP ↑
\\
\colrule
GCN & 2 & N/A &$501,154$ & $ 0.4077 \pm 0.0009$ & $500,456$ & $ 0.4396 \pm 0.0048$
\\
GCN from Ref.~\cite{dwivedi2022long} & 2 & N/A  & $\approx509,000$ & $ 0.3950\pm 0.0017 $ & $\approx509,000$ & $ 0.4566 \pm 0.0059 $
\\
GCN+LapPE & 2 & 4 & $501,066
$ & $ 0.3195 \pm 0.0014$ & $500,372$ & $0.4776\pm0.0040$ 
\\
GCN+PhotPE & 2 & 4 & $501,066$ & $ \mathbf{0.2995\pm 0.0005}$ & $500,372$ & $ \mathbf{0.4886\pm 0.0052}$
\\
\colrule
GCN  & 3 & N/A & $501,446$ & $ 0.4054\pm0.0012 $ &  $500,950$ & $0.4929\pm 0.0066$
\\
GCN+LapPE  & 3 & 4 & $501,375
$ & $ 0.3040\pm 0.0009$ & $500,882$ & $0.4857\pm0.0027$  
\\
GCN+PhotPE  & 3 & 4 & $501,375$ & $ \mathbf{0.2969\pm 0.0016}$ & $500,882$ & $\mathbf{0.5123\pm 0.0011}$
\end{tabular}
\end{ruledtabular}
\end{table*}

\textit{Photonic Positional Embeddings.---}To address the limitations of purely spectral approaches, we introduce PEs derived from photonic dynamics simulation. We model the molecular graph as a network of coupled optical resonators, where a coupling matrix $J$ follows the graph adjacency structure. The system evolution is governed by $d\psi/dt = -\gamma\psi - i J\psi + P_m$, where $\psi(t)$ represents a vector complex field amplitudes, $\gamma$ is the damping rate, and $P_m$ denotes optical pumping at node $m$ (also detailed in Methods). By simulating wave propagation through this photonic network and reading out time-averaged intensities, we obtain a correlation matrix $G$ formed by intensity-intensity correlations $\{G_{ij}\}$ between nodes $i$ and $j$. This matrix captures how excitations from different source nodes overlap throughout the graph (see Methods for more details). The eigenvectors of this correlation matrix serve as PEs $U_{C,k}$, providing each node with features that encode its \emph{global} connectivity patterns. These photonic-inspired embeddings offer a physically-grounded alternative to Laplacian eigenvectors, capturing multi-scale interactions through physical propagation dynamics (similar to coherent walks on graphs) rather than abstract spectral decomposition. Like LapPE, PhotPE can be integrated at every GCN layer through concatenation: $\widetilde{H} = [H \Vert \mathcal{P}_C]$, where $\mathcal{P}_C$ denotes the photonic features. This design ensures that global structural information complements local message passing throughout the network depth. The details of PhotPE are shown in Methods.

\textit{Molecular Graph Datasets.---}We evaluate our approach on the Long Range Graph Benchmark (LRGB) peptide datasets~\cite{dwivedi2022long}, which provide ideal test cases for long-range interaction modeling. \texttt{Peptides-struct} dataset contains $15,535$ peptide molecules from the \texttt{SATPdb}~\cite{singh2016satpdb} database with regression targets spanning $11$ structural properties (molecular weight, secondary structure, solvent accessibility, etc.), while \texttt{Peptides-func} focuses on multi-label classification of $10$ functional categories. With average graph sizes of $150.94$ nodes, $307.30$ edges, and average shortest paths of $20.89\pm9.79$, these datasets specifically challenge models to capture interactions beyond local neighborhoods, precisely where our photonic embeddings provide value.
\begin{figure}[b]
\begin{center}
\includegraphics[width=1.0\linewidth]{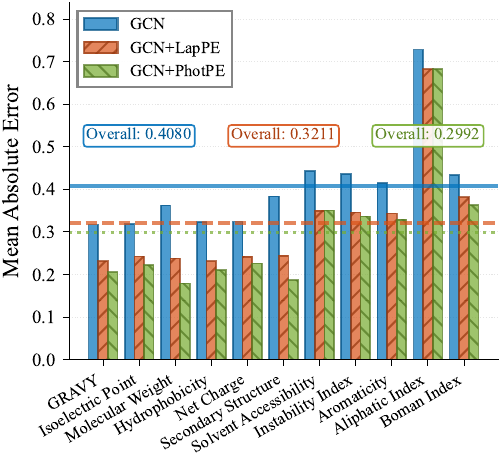}
\end{center}
\vspace{-2\baselineskip}
\caption{\textbf{Regression for different molecular properties.} Mean absolute error (MAE) comparison across $11$ molecular properties in \texttt{Peptides-struct} dataset. Properties include physicochemical descriptors (GRAVY, isoelectric point, molecular weight, hydrophobicity, charge), structural features (secondary structure, solvent accessibility), and stability indices (instability, aromaticity, aliphatic, Boman). The GCN+PhotPE approach (green) consistently outperforms both the baseline GCN (blue) and GCN+LapPE (orange) methods, with horizontal lines indicating overall MAE for each model.}
\label{fig:mae_comparison}
\end{figure}

 
\section{Results}
\label{results}

We proceed to test the proposed pipeline and evaluate how the different PEs perform across molecular property prediction tasks using LRGB datasets~\cite{dwivedi2022long}. First, we quantify the effectiveness of PEs by comparing three model variants: baseline GCN without any positional information, GCN enhanced with Laplacian eigenvector embeddings (GCN+LapPE), and our proposed GCN with photonics-enhanced embeddings (GCN+PhotPE). Our evaluation focuses on two key aspects: (1) the ability to capture long-range dependencies in molecular graphs, as measured by regression accuracy on structural properties, and (2) the capacity to learn complex functional relationships, as assessed through multi-label classification performance. We present results across different network depths (two and three layers), 
aiming to understand how PEs interact with and enrich GCNs. In Section~\ref{experiment}, we discuss possible experimental implementations for getting photonic PEs directly from physical systems.
\begin{figure*}[t]
\begin{center}
\includegraphics[width=1.0\linewidth]{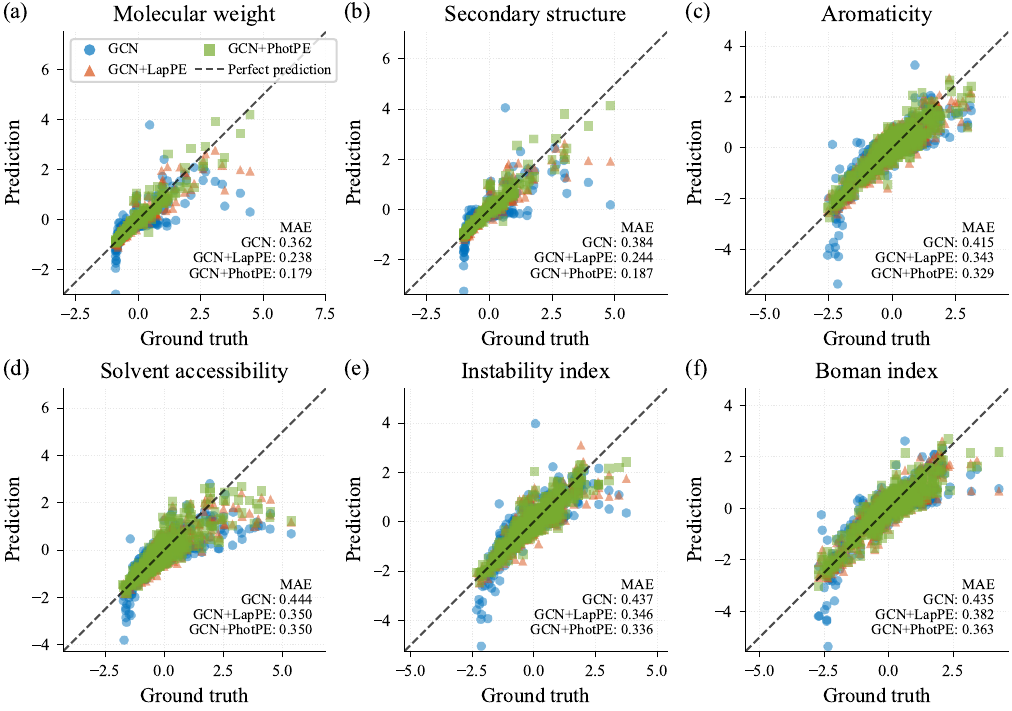}
\end{center}
\vspace{-2\baselineskip}
\caption{\textbf{Molecular property predictions for GCN model variants.} Comparison of ground truth versus predicted values for (a) molecular weight, (b) secondary structure, (c) aromaticity, (d) solvent accessibility, (e) instability index, and (f) Boman index. Three models are shown: GCN (blue), GCN+LapPE (red), and GCN+PhotPE (green). Dashed lines indicate perfect prediction. MAE values are shown in each panel.}
\label{fig:property_predictions}
\end{figure*}


\textit{Regression Tasks.---}\texttt{Peptides-struct} dataset~\cite{dwivedi2022long} is designed as a multi-target graph regression task that predicts 11 aggregated 3D structural properties simultaneously. As detailed in Fig.~\ref{fig:mae_comparison}, these properties span multiple scales of peptide characterization, from atomic-level composition to global physicochemical behavior and functional predictions. The diversity of these properties provides a comprehensive test of the GCN model ability to capture complex structure-property relationships in peptide systems. All properties are normalized to zero mean and unit standard deviation. The task is particularly challenging because it requires implicit understanding of complex 3D molecular interactions without being provided explicit 2D or 3D coordinate information, forcing models to infer spatial relationships purely from the graph structure. For this challenging regression task, GCN+PhotPE achieved remarkable improvements (see Table~\ref{tab:results_main} and Fig.~\ref{fig:mae_comparison}). With $2$ GCN layers, our method attained a test MAE of $0.2995$, representing a $\approx$26.5\% ($\approx$24.2\%) improvement over the baseline GCN (GCN from Ref.~\cite{dwivedi2022long}). The photonics-informed approach consistently outperforms LapPE with achieving a $6.3\%$ and $2.3\%$ reduction in MAE for the $2$ and $3$ layer models, respectively. To visualize these gains, Fig.~\ref{fig:property_predictions} shows representative ground-truth vs prediction plots for several \texttt{Peptides-struct} targets, where GCN+PhotPE yields consistently lower MAE and tighter agreement with the diagonal.

From an application perspective, lower regression error is valuable because many steps in peptide design and screening rely on ranking candidates rather than perfectly calibrated values. A reduced MAE therefore makes it more reliable to prioritize peptides with favorable stability/solubility proxies (e.g. instability, Boman, hydrophobicity-related descriptors) before committing to synthesis.


\textit{Classification Tasks.---}\texttt{Peptides-func} dataset uses the same molecular graphs but focuses on multi-label classification of peptide functions. The task involves predicting $10$ functional classes including antibacterial, antiviral, cell-cell communication, and others~\cite{dwivedi2022long}. The evaluation metric is average precision (AP), calculated as the unweighted mean across all $10$ classes, which is particularly suitable for imbalanced multi-label classification. The results on \texttt{Peptides-func} classification tasks reveal an even more compelling advantage for PhotPE. While the baseline GCN achieves reasonable performance with deeper architectures (AP of $0.4929$ with $3$ layers), the addition of PhotPE boosts performance across all layer configurations. The best performance among $4$-dimensional PE models is achieved by GCN+PhotPE with $3$ layers, yielding an AP of $0.5123$ which represents a $3.9\%$ improvement over the baseline GCN and constitutes a significant improvement over both baseline and Laplacian-enhanced models.

From the perspective of applications, improved classification performance is valuable because peptide screening often proceeds by selecting a small shortlist from a much larger pool of candidates \cite{Vincenzi2024}. In this setting, performance is dominated by how reliably true positives are ranked near the front of the list (early recognition), and even a few percentage points in AP can translate into more functional hits in a fixed-size shortlist, improving the efficiency of subsequent synthesis and validation \cite{Truchon2007EarlyRecognition,Saito2015PR}.

The parameter efficiency of our approach is also noteworthy. All models are constrained to approximately $500\mathrm{K}$ parameters, ensuring fair comparison as suggested by Ref.~\cite{dwivedi2022long}. The best performance of PhotPE is particularly pronounced in shallow networks ($2$ and $3$ layers), where the physics-informed features compensate for the lack of long-range interaction. This suggests our approach could enable more parameter-efficient architectures for large molecular tasks and makes it practical for deployment in resource-constrained settings where model size is a concern. Although introducing additional PEs information can greatly help the model generalize faster, it can also cause overfitting, especially with deeper GCN layers (see Table~\ref{tab:extended_results} in Appendices for details).


\section{Experimental considerations}
\label{experiment}

To implement the photonic correlation-based PEs, we build upon programmable synthetic-frequency optical lattices realized in a fiber ring cavity~\cite{Yang2022,Yuan2018,senanian2023programmable}. In this platform, each graph node corresponds to a discrete cavity frequency mode spaced by the cavity free spectral range (FSR), which we denote as $\Omega$. For the system in Ref.~\cite{senanian2023programmable} with $\Omega = 1.226\,\mathrm{MHz}$, the achievable graph size is primarily set by the usable optical bandwidth over which frequency modes can circulate with sufficient signal-to-noise ratio. An intracavity optical bandpass filter of bandwidth $\Delta f$ can select a block of cavity modes, such that the total number of available nodes is approximately $N \approx \Delta f/\Omega$. With a $\Delta f \approx 125\,\mathrm{GHz}$ filter (corresponding to $\sim 1\,\mathrm{nm}$ around $1550\,\mathrm{nm}$ wavelength), this yields $N \sim 10^{5}$ frequency modes. Ref.~\cite{senanian2023programmable} has demonstrated spectral measurements with graphs sizes over $100,000$ nodes. Larger graphs can be considered if a wider low-noise optical bandwidth is supported (e.g. by increasing $\Delta f$ while suppressing amplified spontaneous emission and maintaining gain and dispersion conditions).

Programmable graph connectivity is implemented by an electro-optic phase modulator (EOM) driven by a multi-tone radio-frequency waveform (Fig.~\ref{fig:workflow}c). The modulator generates controlled sidebands that couple cavity modes separated by integer multiples of the FSR, $i \leftrightarrow j = i \pm m$, generating tunable hoppings $J_{ij}$ (adjacency) between nodes in the synthetic-frequency lattice. The achievable coupling range is determined by the highest drive frequency that can be applied with sizable modulation depth. With a $40\,\mathrm{GHz}$ phase modulator, the maximum hopping index can be estimated as $m_{\max} \approx 40\,\mathrm{GHz}/\Omega \approx 3\times 10^{4}$, allowing long-range couplings over tens of thousands of mode steps. In practice however the usable coupling range at a given strength can be constrained by modulation efficiency and RF power.

The readout of the system can be performed in parallel via heterodyne detection by mixing the output with a frequency-shifted local oscillator on a $30\,\mathrm{GHz}$ photodetector and analysed in the RF domain. With an RF acquisition bandwidth $\beta_{\mathrm{RF}}\approx 26\,\mathrm{GHz}$ and mode spacing $\Omega=1.226\,\mathrm{MHz}$, this enables simultaneous, site-resolved readout of blocks of size $N_{\parallel}\approx \beta_{\mathrm{RF}}/\Omega \approx 2\times 10^{4}$ nodes per measurement window, and the full $N\sim 10^{5}$ graph can be covered by stitching together five of such blocks. The intrinsic dynamical timescale of the photonic processor is set by the cavity roundtrip time $\tau_{\mathrm{rt}}=1/\Omega \approx 0.82\,\mu\mathrm{s}$, so the fastest update/propagation rates are naturally on the order of $\tau_{\mathrm{rt}}$. In practice, total measurement time is set by transient settling, readout tiling overheads, and the chosen integration time for adequate signal-to-noise. These timescales have to be compared with the timescales for LapPE-based approach, taking $0.011$ seconds per sample ($166$\,s for the full dataset) when ran on GPU (NVIDIA GeForce RTX 4090). We believe that physical accelerators for PhotPE can potentially help reduce both training and inference times, as well as the energy cost of running GCN+PE models.


\section{Discussion and Conclusions}
\label{discussion_and_conclusions}

We have introduced a physics-enhanced framework that draws inspiration from photonic wave dynamics to enhance graph learning. By embedding photonic correlations into GCNs, we demonstrated improved performance on challenging long-range graph benchmarks, highlighting the potential of physical inductive biases to overcome structural limitations of conventional message-passing architectures. Our key contribution lies in bridging optical computing principles with graph ML through synthetic frequency lattice simulations.

Beyond algorithmic gains, this work points to an emerging direction where photonic devices are used not only as a source of inspiration, but also serve as a platform for accelerating ML. The same principles of wave propagation that enrich graph embeddings \textit{in silico} could be harnessed in photonic hardware, enabling direct analog computation of correlations that are otherwise costly to simulate digitally. Such an approach opens avenues toward scalable, energy-efficient photonic architectures \cite{McMahon2023}.

Regarding future directions, an intriguing possibility appears for optical lattices that can experience non-Abelian gauge fields~\cite{Cheng2025,ChengFan2023}. Another opportunity is to use programmable quantum synthetic lattices \cite{Bartlett2024} that include nonlinear operation, which can be implemented with quantum nonlinear nodes \cite{Munoz-Matutano2019,Delteil2019,Kuriakose2022}.


\begin{acknowledgments}
    O.\,K. thanks Peter L. McMahon for useful discussions on the subject. The authors acknowledge the support from UK EPSRC grant EP/X017222/1. O.\,K. also acknowledges support from UK EPSRC EP/Z53318X/1.
\end{acknowledgments}


\section*{Methods}
\label{methods}

\subsection{Laplacian-based positional embeddings}

A common way to equip each node with a positional feature is to use the eigenvectors of the graph Laplacian. The symmetric normalized Laplacian is defined as
\begin{equation}
L_{\text{sym}} = \mathbb{I} - D^{-1/2} A D^{-1/2},
\end{equation}
which admits the eigendecomposition
\begin{equation}
L_{\text{sym}} = U_L \Lambda_L U_L^{\top},
\end{equation}
where $U_L = [u_{L,1}, u_{L,2}, \ldots, u_{L,n}]$ contains orthonormal eigenvectors and $\Lambda_L = \mathrm{diag}(\lambda_{L,1}, \ldots, \lambda_{L,n})$ with eigenvalues $0 = \lambda_{L,1} \leq \lambda_{L,2} \leq \cdots \leq \lambda_{L,n} \leq 2$.
\begin{figure*}[t]
\begin{center}
\includegraphics[width=1.0\linewidth]{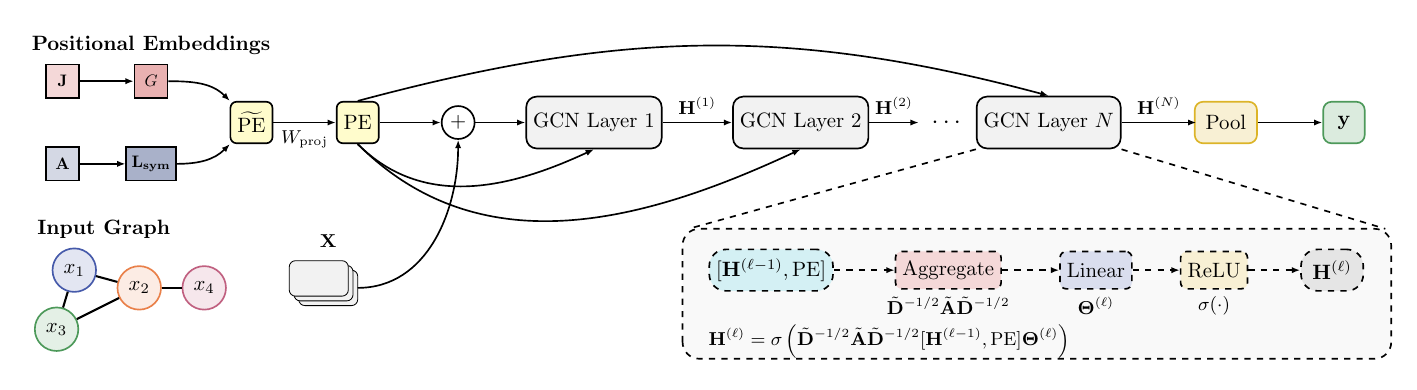}
\end{center}
\vspace{-2\baselineskip}
\caption{\textbf{Architecture of the graph convolutional network with Laplacian and photonic-enhanced positional embeddings (PEs).} The model takes an input graph $X$ with nodes ($x_{1}, x_{2}, x_{3}, x_{4}$) and processes it through two parallel pathways to generate PEs. The photonic-enhanced PE (taking coupling matrix $J$ to the correlation matrix $G$) and graph Laplacian PE ($A$, $L_\mathrm{sym}$) are combined via a projection matrix $W_\mathrm{proj}$ to create PE. These PEs are concatenated with the node features and fed into a stack of $N$ GCN layers. Each GCN layer (detailed in the bottom inset) performs graph convolution by: concatenating the previous layer's hidden representation $H^{(\ell-1)}$ with PE, applying the normalized adjacency matrix aggregation $\widetilde{D}^{(-1/2)}\widetilde{A}\widetilde{D}^{(-1/2)}$, linear transformation with parameters $\Theta^{(\ell)}$, and ReLU activation $\sigma(\cdot)$. The final layer output $H^{(N)}$ is pooled to produce the prediction $y$. This architecture uses structural (Laplacian) and physical (photonic) properties of the graph to enhance node representation learning.}
\label{fig:architecture}
\end{figure*}

We construct $k$-dimensional PEs using the first $k$ non-trivial eigenvectors~\cite{dwivedi2022long}
\begin{equation}
U_{L,k} = [u_{L,2} \| u_{L,3} \| \cdots \| u_{L,k+1}] \in \mathbb{R}^{n \times k}.
\end{equation}
To ensure stability across different graphs, each eigenvector is normalized to zero mean and unit variance,
\begin{equation}
\tilde{u}_{L,i} = \frac{u_{L,i} - \mu(u_{L,i})}{\Sigma(u_{L,i})},
\label{eq:normalization_eigenvector}
\end{equation}
where $\mu(\cdot)$ and $\Sigma(\cdot)$ denote the mean and standard deviation across nodes. Thus, we have $\widetilde{U}_{L,k} = [\tilde{u}_{L,2} \Vert \tilde{u}_{L,3} \Vert\cdots\Vert \tilde{u}_{L,k+1}] \in\mathbb{R}^{n\times k}$. These $\widetilde{U}_{L,k}$ can be concatenated via linear projection with dimension $d_{\text{proj}}$ to any existing node feature $X\in\mathbb{R}^{n\times d}$ to yield
\begin{equation}
\mathcal{P}_L = [X \Vert W_{\text{proj}}\widetilde{U}_{L,k}]\in\mathbb{R}^{n\times (d+d_{\text{proj}})}.
\end{equation}
Such Laplacian eigenfeatures have been shown to improve expressivity in GNNs by providing a global feature on the graph \cite{dwivedi2023benchmarking}. Unlike approaches that only prepend PEs to input features, we inject the projected PE at every GCN layer: $\widetilde{H}^{(\ell)} = [H^{(\ell)} \Vert W_{\text{proj}} \widetilde{U}_{L,k}]$, where $W_{\text{proj}}$ is a shared learnable projection matrix. This design choice allows the model to use global structural information throughout the network depth. Note that the first GCN layer ($H^{(0)}$) uses the raw data with the LapPE. Thus, the injection of the LapPE for each GCN layer is given by
\begin{equation}
\label{eq:projection_LPE}
\widetilde{H}^{(\ell)} = [H^{(\ell)} \Vert \mathcal{P}_L].
\end{equation}
\begin{figure*}
\begin{center}
\includegraphics[width=0.8\linewidth]{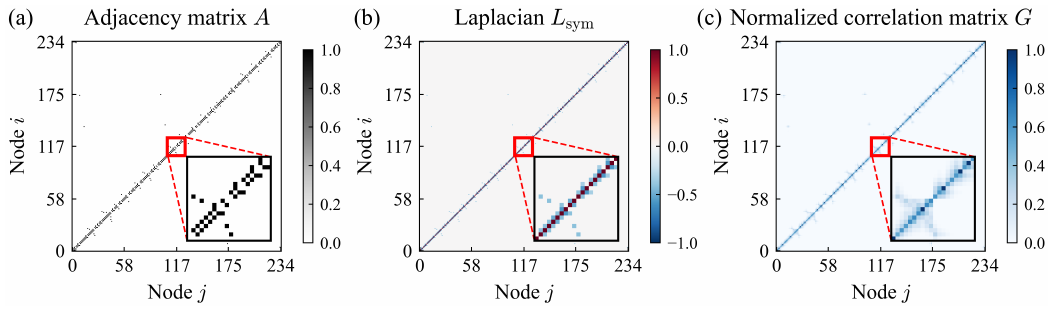}
\end{center}
\vspace{-2\baselineskip}
\caption{\textbf{Representations of peptide molecular graph structure.} (a) Adjacency matrix $A$ showing the binary connectivity pattern between nodes (heavy atoms), where matrix elements are $1$ for connected nodes and $0$ otherwise. (b) Graph Laplacian $L_{\mathrm{sym}} = \mathbb{I} - D^{-1/2}AD^{-1/2}$, capturing the spectral properties of the graph structure. (c) Normalized correlation matrix $G$ computed from photonic dynamics. Matrix dimensions correspond to the number of nodes ($235$) in the peptide structure ($235\times235$). The insets (red box) in each panel highlight part of the diagonal region.}
\label{fig:matrices}
\end{figure*}

\subsection{Correlation-based positional embeddings from photonic dynamics}

We now introduce our approach based on photonic dynamics simulation. Consider a photonic system where nodes represent optical resonators coupled according to the graph structure \cite{senanian2023programmable,Jacqmin2014,Whittaker2018}. The dynamics of complex field amplitudes $\psi(t) \in \mathbb{C}^n$ follow
\begin{equation}\label{eq:photonic_dynamics}
\frac{d\psi}{dt} = -\gamma\psi - i J\psi + P_m,
\end{equation}
where $J = A/\max(A)$ is the normalized coupling matrix, $P_m$ is the $m$-th injection, and $\gamma$ is the damping rate. Note that in the case of unweighted graphs as we consider in this work, $A$ can be directly used as the coupling matrix. From a ML perspective, this dynamics serves as a learnable feature extractor that captures graph topology through physical simulation. The damping parameter $\gamma$ acts as a regularizer, controlling the locality of correlations, while the evolution time $T$ determines the receptive field of our features. 

Solving Eq.~\eqref{eq:photonic_dynamics} with constant pumping $P_m$ over $t\in[0,T]$ yields $\psi(t)$. We define the time-averaged intensity as
\begin{equation}
v_{m}(j) = \frac{1}{T-t_s}\int_{t_s}^T |\psi_j(t)|^2 dt,
\end{equation}
where $t_s$ marks the starting time and $T$ is the total evolution time. Collecting all pairs $(m,j)$ forms the matrix $V\in\mathbb{R}^{n\times n}$. The correlation between nodes $i,j$ is then obtained by
\begin{equation}\label{eq:Gij}
G_{ij} = \frac{1}{n}\sum_{m=1}^n v_m(i)v_m(j) - \overline{v}(i)\overline{v}(j),
\end{equation}
in which the mean intensity $\overline{v}$ is defined by $\overline{v}(j)=n^{-1}\sum_mv_m(j)$. This correlation matrix encodes how excitations from different source nodes overlap at pairs of target nodes, capturing global connectivity patterns. To facilitate the machine learning workflow, we normalize $G$ to $[0,1]$.

We compute the first-$k$ eigenvectors of $G$ by solving $G u_C = \lambda u_C$ and ordering by descending $\lambda$. 
We denote the resulting matrix of node features as $U_{C,k}\in\mathbb{R}^{n\times k}$. Similar to the treatment of Eq.~\eqref{eq:projection_LPE}, we have the expression of the GCN layer
\begin{equation}
\widetilde{H} = [H \Vert \mathcal{P}_C],
\end{equation}
with injection of PhotPE being
\begin{equation}
\mathcal{P}_C = [X \Vert W_{\text{proj}}\widetilde{U}_{C,k}]\in\mathbb{R}^{n\times (d+d_{\text{proj}})}.
\end{equation}
Note that $\widetilde{U}_{C,k} = [\tilde{u}_{C,2} \Vert \tilde{u}_{C,3} \Vert\cdots\Vert \tilde{u}_{C,k+1}] \in\mathbb{R}^{n\times k}$ is obtained through the same normalization introduced at Eq.~\eqref{eq:normalization_eigenvector} for the Laplacian case.

\subsection{Implementation and simulation setup}

We implement three model variants: baseline GCN, GCN with LapPE, and GCN with PhotPE. All models share identical architectures with $2$ and $3$ GCN layers (see Table~\ref{tab:extended_results} for results with $4$ and $5$ layers) and hidden dimensions tuned to maintain approximately $500\mathrm{K}$ parameters. For PhotPE computation, we use $T=2.0$, $\gamma=0.5$, and $1000$ integration steps, with eigenvectors extracted from the correlation matrix after normalizing to $[0,1]$. To obtain the correlation matrix, we average the node intensities over the final $99\%$ of the trajectory from the numerical solution to Eq.~\eqref{eq:photonic_dynamics}, ensuring proper propagation of the photonic excitation throughout the graph while retaining additional interactions present in the correlation matrix. These parameters are fixed across all datasets to ensure consistency. The PE dimension is set to $4$ for all the chosen layers (set to $2$ for the extra 4-layers case, see Table~\ref{tab:extended_results}) with an equivalent projection dimension.

For both Peptides tasks, we use the Adam optimizer with batch size $128$ and implement learning rate scheduling via ReduceLROnPlateau with factor $0.5$ and patience $20$. The framework supports grid search over learning rates ranging from $0.3\times10^{-3}$ to $10^{-3}$ with each step being $10^{-4}$, which is similar to the implementation of machine-learning training in Ref.~\cite{dwivedi2022long}. Early stopping is triggered when the learning rate falls below $10^{-5}$ or validation performance plateaus for $50$ epochs. All experiments are executed with $4$ random seeds to ensure statistical robustness.

\subsection{Photonic dynamics parameters}

For the photonic correlation computation, we set the final time $T = 2.0$ and the damping coefficient $\gamma = 0.5$. To the numerical solution to Eq.~\eqref{eq:photonic_dynamics}, 1000 Euler integration steps with time step $\Delta t = 0.002$ are used. The steady-state regime begins at $t_s=0.02$ (manually set), which means that we average the node intensities over the final $99\%$ of the trajectory. These parameters were fixed across all samples to ensure consistency.

\subsection{Dataset details}

Both datasets (\texttt{Peptides-struct} and \texttt{Peptides-func}) employ identical data splits to ensure consistency: $10,873$ samples ($70\%$) for training, $2,331$ samples ($15\%$) for validation, and $2,331$ samples ($15\%$) for testing. The splits are computed using stratified splitting based on meta-classes to maintain balanced distributions. For \texttt{Peptides-struct}, performance is evaluated using mean absolute error (MAE) as the primary metric across all properties. For \texttt{Peptides-func}, the evaluation metric is average precision (AP), calculated as the unweighted mean across all $10$ classes, which is particularly suitable for imbalanced multi-label classification.

These datasets are specifically designed to benchmark the ability of GNNs to capture long-range interactions, and are particularly well-suited for evaluating positional embeddings. Unlike ZINC dataset \cite{ZINCdataset} where the small graph sizes ($<40$ nodes) may cause information over-squashing effects to be negligible, the Peptides datasets with large graph sizes require models to propagate information well beyond local neighborhoods. This makes them ideal for evaluating whether positional embeddings can effectively encode global structural information and help graph architectures handle long-range dependencies.

\nocite{*}

\appendix


\begin{table*}[t]
\caption{\label{tab:extended_results}Baselines for \texttt{Peptides-struct} and \texttt{Peptides-func} with deeper architectures. Performance metric is mean absolute error (MAE, lower is better) and average precision (AP, higher is better) for \texttt{Peptides-struct} and \texttt{Peptides-func} tasks, respectively. Each experiment was run with 4 different seeds with 4 and 5 GCN layers. N/A: Not applicable. \textbf{Bold}: Best score.}
\begin{ruledtabular}
\begin{tabular}{c c c c c c c}
& &
 & \multicolumn{2}{c}{\text{Peptides-struct} (Regression)} & \multicolumn{2}{c}{\text{Peptides-func} (Classification)}
\\
\textrm{Model} & 
\# \textrm{Layers}&
\# \textrm{Dim. PE}&\# \textrm{Parameters} & Test MAE ↓
& \# \textrm{Parameters} & Test AP ↑
\\
\colrule
GCN  & 4 & N/A & $501,806$ & $ 0.3835\pm 0.0059$ &  $501,400$ & $0.5262\pm0.0087$
\\
GCN+LapPE  & 4 & 2 & $500,138
$ & $0.3001\pm 0.0033$ & $499,734$ & $0.5078\pm0.0040$  
\\
GCN+PhotPE  & 4 & 2 & $500,138$ & $ 0.3014\pm 0.0026$ & $499,734$ & $\mathbf{0.5386\pm 0.0033}$
\\
GCN+LapPE  & 4 & 4 & $500,919
$ & $ 0.2985\pm 0.0010$ & $500,516$ & $0.4839\pm0.0042$  
\\
GCN+PhotPE  & 4 & 4 & $500,919$ & $ \mathbf{0.2936\pm 0.0011}$ & $500,516$ & $0.5239\pm 0.0041$
\\
\colrule
GCN  & 5 & N/A & $501,590$ & $ 0.3693\pm 0.0027$ &  $501,238$ & $0.5325\pm 0.0077$
\\
GCN from Ref.~\cite{dwivedi2022long}  & 5 & N/A  & $\approx508,000$ & $0.3496\pm0.0013$ & $\approx508,000$ & $\mathbf{0.5930\pm0.0023}$ 
\\
GCN+LapPE  & 5 & 4 & $500,103
$ & $ 0.2948\pm 0.0025$ & $499,754$ & $0.4878\pm 0.0051$  
\\
GCN+PhotPE  & 5 & 4 & $500,103$ & $ \mathbf{0.2929\pm 0.0026}$ & $499,754$ & $0.5236\pm 0.0044$
\end{tabular}
\end{ruledtabular}
\end{table*}

\section{Preliminaries on Graph Convolutional Networks}
\label{sec:Preliminaries_on_GCN}

The graph Laplacian is defined by
\begin{equation}
L = D - A,
\end{equation}
and its symmetric normalization is given by
\begin{equation}
L_{\mathrm{sym}} = D^{-1/2} L D^{-1/2} = \mathbb{I} - D^{-1/2} A D^{-1/2}.
\end{equation}
Since $L_{\mathrm{sym}}$ is real symmetric and positive semidefinite, it admits an eigendecomposition
\begin{equation}
L_{\mathrm{sym}} = U \Lambda U^\top,
\end{equation}
where $U=[u_1,\ldots,u_n]$ is orthonormal, and $\Lambda = \mathrm{diag}(\lambda_1,\dots,\lambda_n)$ collects the eigenvalues $0=\lambda_1\le\cdots\le\lambda_n$.
\begin{figure*}[t]
\begin{center}
\includegraphics[scale=0.8]{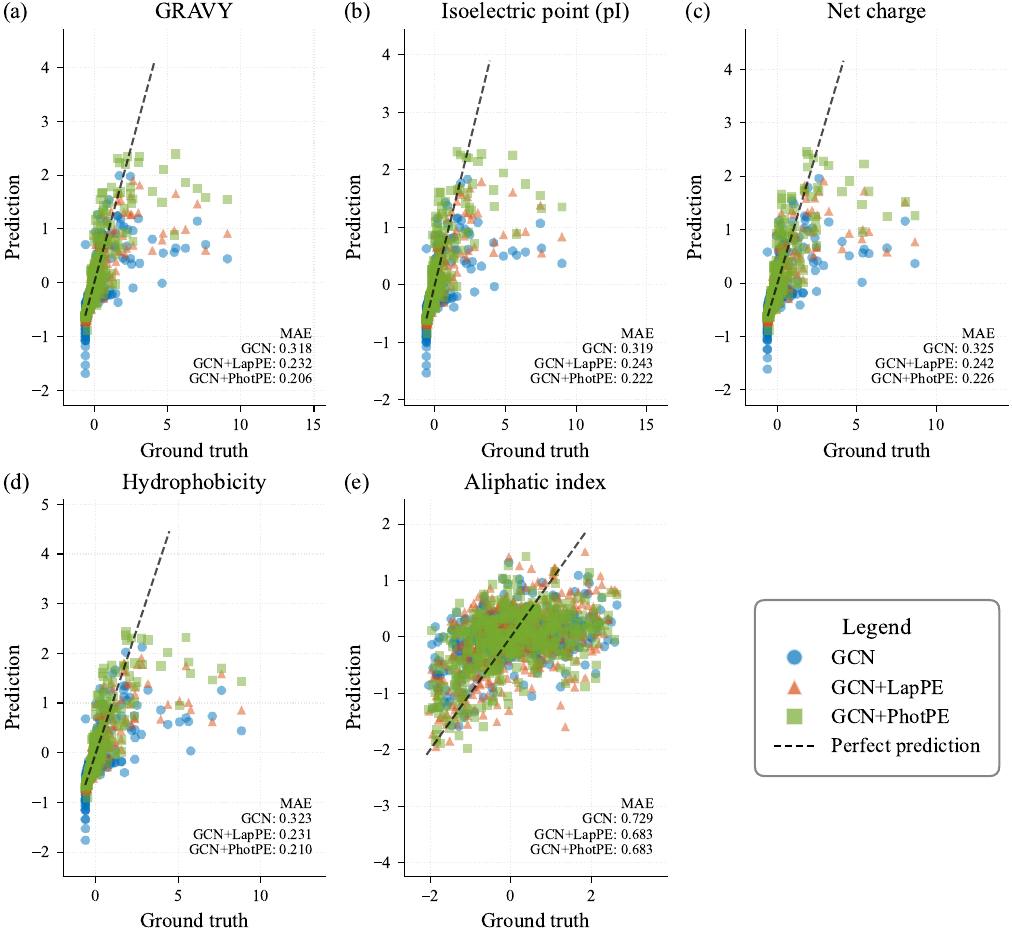}
\end{center}
\vspace{-2\baselineskip}
\caption{\textbf{Additional molecular property predictions for GCN model variants.} Comparison of ground truth versus predicted values for (a) GRAVY, (b) isoelectric point, (c) charge, (d) hydrophobicity, and (e) aliphatic index. Three models are shown: GCN (blue), GCN+LapPE (red), and GCN+PhotPE (green). Dashed lines indicate perfect prediction. MAE values are shown in each panel.}
\label{fig:appendix_properties}
\end{figure*}
Given a graph signal $X\in\mathbb{R}^n$ assigning a scalar to each node, its graph Fourier transform is defined by
\begin{equation}
\widehat{X} = U^\top X,
\end{equation}
and the inverse transform recovers $X=U\widehat{X}$. The columns of $U$ play the role of Fourier modes, and the corresponding eigenvalues play the role of frequencies. Thus, we can define the graph filter $\mathcal{G}$, which can be specified by its spectral response
\begin{equation}
\widehat{\mathcal{G}}(\Lambda)=\mathrm{diag}(\widehat{\mathcal{G}}(\lambda_1),\dots,\widehat{\mathcal{G}}(\lambda_n)).
\end{equation}
The convolution of $\widehat{\mathcal{G}}$ with a signal $X$ is defined as
\begin{equation}
\label{eq:graph_convolutional_kernel}
\widehat{\mathcal{G}} \ast X = U \,\widehat{\mathcal{G}}(\Lambda)\, U^\top X.
\end{equation}
However, directly implementing Eq.~\eqref{eq:graph_convolutional_kernel} is expensive for large graphs due to the full eigendecomposition. To reduce the computational cost, one often replaces $\widehat{\mathcal{G}}(\Lambda)$ by a $K$-th order polynomial in $\Lambda$,
\begin{equation}\label{eq:poly_filter}
\widehat{\mathcal{G}} \approx \sum_{k=0}^K \theta_k \Lambda^k.
\end{equation}
Substituting Eq.~\eqref{eq:poly_filter} into Eq.~\eqref{eq:graph_convolutional_kernel} yields an efficient form
\begin{align}
\label{eq:graph_convolutional_kernel_polynomial_form}
\widehat{\mathcal{G}} \ast X &\approx U \Big (\sum_{k=0}^K \theta_k \Lambda^k\Big ) U^\top X
\\
&= \sum_{k=0}^K \theta_k \, (U \Lambda^k U^\top) X
\\
&= \sum_{k=0}^K \theta_k \, L_{\mathrm{sym}}^k \; X.
\end{align}
This avoids explicit spectral decomposition and relies instead on repeated sparse multiplications by $L_{\mathrm{sym}}$.

Setting $K=1$, coefficients $\theta_0=2\theta$ and $\theta_1=-\theta$ in Eq.~\eqref{eq:graph_convolutional_kernel_polynomial_form} yields the GCN convolutional kernel
\begin{equation}\label{eq:gcn_kernel}
\widehat{\mathcal{G}} \ast X = \theta\big(\mathbb{I} + D^{-1/2}AD^{-1/2}\big) X.
\end{equation}
Introducing the \emph{renormalization trick} by adding self-loops, we can define $\tilde A = A + \mathbb{I}$, $\tilde D = D + \mathbb{I}$, and replace $\mathbb{I} + D^{-1/2}AD^{-1/2}$ with $\tilde D^{-1/2}\tilde A\tilde D^{-1/2}$. Thus, we have
\begin{equation}\label{eq:gcn_kernel_renorm_trick}
\widehat{\mathcal{G}} \ast X = \theta\big(\tilde D^{-1/2}\tilde A\tilde D^{-1/2}\big) X.
\end{equation}

\clearpage

\newpage


%

\end{document}